\documentclass[fleqn,10pt]{wlscirep}
\usepackage[utf8]{inputenc}
\usepackage[T1]{fontenc}
\usepackage[normalem]{ulem}
\usepackage{multirow}
\usepackage{lineno}

\newcommand{\blue}[1]{{\color{blue}#1}}

\title{
Room-temperature polaron-mediated
polariton nonlinearity in
MAPbBr$_3$ perovskites \\
}

\author[1,$^{\dagger}$]{M.A.~Masharin}
\author[1,$^{\dagger}$,*]{V.A.~Shahnazaryan}
\author[1]{I.V.~Iorsh}
\author[1,2]{S.V.~Makarov}
\author[1,3]{A.K.~Samusev}
\author[1,4]{I.A.~Shelykh}

\affil[1]{ITMO University, School of Physics and Engineering, St. Petersburg, 197101, Russia}
\affil[2]{Qingdao Innovation and Development Center, Harbin Engineering University, Qingdao 266000, Shandong, China}
\affil[3]{Experimentelle Physik 2, Technische Universit\"at Dortmund, 44227 Dortmund, Germany}
\affil[4]{Science Institute, University of Iceland, Dunhagi 3, IS-107, Reykjavik, Iceland}
\affil[*]{Corresponding author}
\affil[$^{\dagger}$]{These authors contributed equally to this work}

\begin{abstract}
Systems supporting exciton-polaritons represent solid-state optical platforms with a strong built-in optical nonlinearity provided by exciton-exciton interactions.
In conventional semiconductors with hydrogen-like excitons the nonlinearity rate demonstrates the inverse scaling with the  binding energy.  This  makes excitons stable at room temperatures weakly interacting, which obviously limits the possibilities of practical applications of the corresponding materials for nonlinear photonics.
We demonstrate experimentally and theoretically, that these limitations can be substantially softened in hybrid perovskites, such as MAPbBr$_3$ due to the crucial role of the polaron effects mediating the inter-particle interactions.
The resulting exciton-polaron-polaritons remain both stable and strongly interacting at room  temperature, which is confirmed by large nonlinear blueshifts of lower polariton branch energy under resonant femtosecond laser pulse excitation. Our findings open novel perspectives for the management of the exciton-polariton nonlinearities in ambient conditions.  
\end{abstract}

\makeatother

\begin{document}

\flushbottom
\maketitle

\thispagestyle{empty}

Excitons are bound states of an electron in a conduction band and a hole in a valence band holding together due to the Coulomb attraction. In direct bandgap materials they can be created by the absorption of photons and annihilate with the photon emission. In the specific designs, when a material with an optical excitonic transition is embedded inside an optical cavity, and the energies of a cavity mode and excitonic transition are brought close to resonance, the regime of the strong light-matter coupling can be achieved \cite{Kavokin2017_OxfPr}. This happens when the characteristic coupling strength, defined by the optical dipole moment of the excitonic transition, exceeds all characteristic broadenings.  Strong coupling has been demonstrated for a variety of experimental geometries including planar Bragg microcavities\cite{weisbuch1992observation}, planar waveguides\cite{ciers2017propagating}, or photonic crystal slabs supporting photonic bound states in continuum (BIC) and leaky modes \cite{zhang2018photonic, kravtsov2020nonlinear}. 

Hybridization between excitons and cavity photons results in  appearance of a new type of the composite elementary excitations in the system, known as exciton-polaritons. The combination of the extremely small effective mass of the polaritons (about five orders of magnitude smaller compared to free electron mass) with macroscopically large coherence length \cite{Ballarini2017} and strong nonlinear optical response provided by exciton-exciton interactions \cite{ciuti1998role,glazov2009polariton} make polariton systems attractive candidates for realization of the nonlinear optical elements of new generation \cite{liew2011polaritonic}, including the extra low threshold lasers \cite{kasprzak2006bose} and all-optical integrated circuits \cite{Amo2010_NatPhot,Askitopoulos2018,Chen2022,Nigro2022}. Moreover, polaritons present an ideal platform for study of collective quantum phenomena at surprisingly high temperatures \cite{carusotto2013quantum}. 

However, for practical applications of polaritonics there is still a gap in the material base. Indeed, the use of the structures with quantum wells based on conventional semiconductor materials, such as GaAs\cite{bajoni2008polariton,nguyen2013realization} and CdTe\cite{kasprzak2006bose} are limited to cryogenic temperatures. On the other hand,  production of the high quality samples based on wide gap materials such as GaN\cite{semond2005strong,liu2015strong} or ZnO \cite{van2006exciton} require expensive fabrication methods,  transition metal dichalcogenide-based systems \cite{kravtsov2020nonlinear,zhao2021ultralow} (TMDs) are difficult to scale up and polymer materials (MeLPPP\cite{zasedatelev2019room}, mCherry\cite{betzold2019coherence}) are highly disordered and are still mostly used in bulky vertical Bragg cavities, which are barely compatible with on-chip designs. Moreover, as all these materials possess tightly bound excitons, which means that the exciton-exciton interaction, and consequently the nonlinear optical coefficient are substantially reduced.

In this context, halide perovskites are very promising candidates for practical applications of polaritonics. This is due to the combination of their remarkable properties, namely stability of excitons at room temperatures, high optical oscillator strength of the excitonic transitions,  defect tolerance, availability of well developed and cheap fabrication and nanostructuring and simple scaling-up technologies~\cite{su2018room,ArXiv2022}, and the ability to support high-Q optical modes at the nanoscale due the high refractive index of the material\cite{su2021perovskite}. The latter allows realizing a planar optical cavity in the geometry of a photonic crystal slab (PCS) made directly from the active excitonic material with high localization of optical field and thus enhanced light-matter interaction. Moreover, it is well known that due to the polar nature of their crystalline lattice, hybrid halide perovskites posses strong electron-phonon interaction\cite{wright2016}, which leads to the substantial enhancement of their optical nonlinear properties, as we will show below.

The nonlinear optical response of exciton-polaritons is determined by their excitonic component and can be experimentally detected as a blueshift of the lower polariton line with increase of the intensity of the external pump. This phenomenon is mainly due to the two mechanisms. The first one is related to the exciton-exciton  interactions, for which exchange contribution is dominant \cite{ciuti1998role,glazov2009polariton} and which apparently increases with an increase of the exciton concentration. The second one is connected with the saturation of the optical absorption related to the composite quantum statistics of excitons, and results in the quenching of the Rabi splitting. The combination of these two effects effects allows to successfully use polariton systems in nonlinear photonic devices\cite{sanvitto2016road} with characteristic response times up down to the sub-picosecond range \cite{chen2022optically} with perovskites being one of the promising material platforms. Polariton nonlinearity has been already demonstrated experimentally in all-inorganic and so-called quasi-2D perovskites  \cite{su2018room,su2017room,fieramosca2019two}, but corresponding properties of hybrid organic-inorganic perovskites (HOIPs) were not yet investigated.  Hybrid perovskites possess certain important differences from  all-inorganic perovskites, such as soft polar crystal lattice favorable for  polaron formation\cite{miyata2017large}. Importantly, polaron effects can be expected to modify the interaction potential \cite{soufiani2015polaronic}, which leads to the enhancement of the exciton binding energy and substantial increase of the polariton nonlinear response \cite{masharin2022polaron}.

Among HOIPs bromide-based perovskite MAPbBr$_3$ is a perspective material for room-temperature polaritonics, since it has an exciton with sufficiently high binding energy and high oscillator strength stable even in the  polycrystalline phase\cite{soufiani2015polaronic,shi2020exciton}.  Previously, strong light-matter coupling regime was reported in MAPbBr$_3$ nanowires \cite{shang2018surface} and thin films embedded in  vertical Bragg cavities \cite{bouteyre2019room}.  Polariton optical nonlinearity was not studied in both mentioned geometries so far, but was investigated in the related material MAPbI$_3$ \cite{masharin2022polaron}, where record high values of the polariton blueshift (up to 19.7 meV) were reported. However, relatively low exciton binding energy in MAPbI$_3$ makes the observed effect possible only at cryogenic temperatures. This obviously limits the possibility of the practical applications, and the study of the nonlinear response of MAPbBr$_3$ thus represents an important contribution in the field.

In this work, we use MAPbBr$_3$ thin film as an active medium with room-temperature stable excitons. We structure the film using nanoimprint lithography method with a periodic grating mold, to get a photonic crystal slab (PCS). We demonstrate that in this geometry the coupling of an excitonic transition with leaky modes of the PCS results in the formation of robust exciton-polaritons. We study experimentally the temperature dependence of the properties of this system and demonstrate, among the rest, the pronounced polariton nonlinear response at room temperature. Moreover, we present a theoretical model which reveals the main mechanisms of the polariton nonlinearity and gives the quantitative description of the experimental data. Potential perspectives of MAPbBr$_3$ in exciton-polariton applications are also highlighted.

\section*{Sample and experimental setup}

\begin{figure}
\centering
\center{\includegraphics[width=0.8\linewidth]{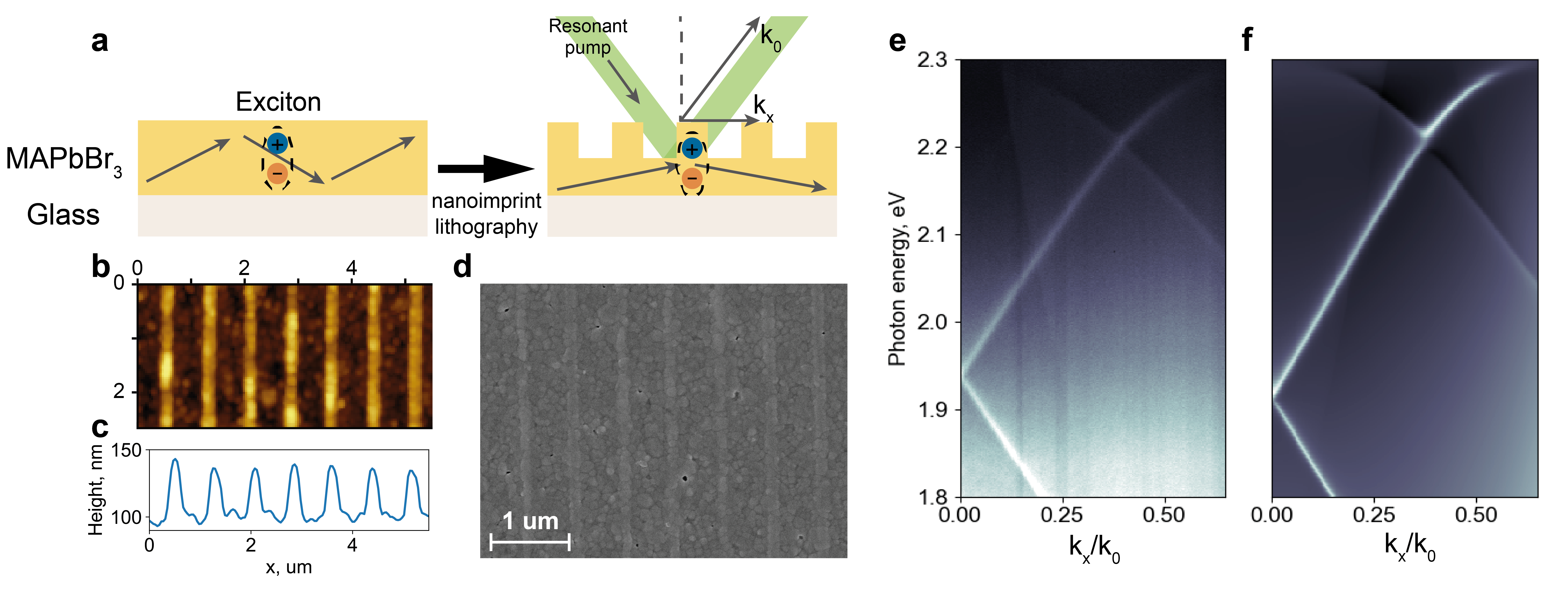}}
\caption{(a) Concept image of a 1D MAPbBr$_3$ PCS fabrication by nanoimprint lithography. In a thin film of MAPbBr$_3$ excitonic transition is strongly coupled with a waveguide photonic mode, which lies below the light cone. The presence of the periodic grating leads to the transformation of the photonic waveguide modes into leaky PCS modes, which makes possible direct optical excitation and detection of the polaritons (b) The morphology of the fabricated 1D MAPbBr$_3$ PCS, measured by AFM method. The color shows the height in each point of the sample from low (dark colors) to high (light colors) (c) The horizontal profile of the studied sample, obtained from the AFM measurement. The data show the presence of the accurate ridges with the height about 100 nm. (d) Image of the studied sample, made by SEM, which shows the grating period of 750 nm and the ridge width of 200 nm. (e) Experimentally measured angle-resolved reflectance spectrum of fabricated MAPbBr$_3$ PCS in TE polarization, which shows the dispersion of the PCS leaky modes. The wavevector of light in free space $k_0$ is defined as $k_0 = E / (\hbar * c)$, where $E$ is photon energy, $c$ is the speed of light. The curvature of the mode near the 2.3 eV is attributed to the coupling with the excitonic transition. In the spectral region above 2.3 eV, there is no photonic mode due to the high optical absorption. (f) The simulated angle-resolved spectrum of the studied structure obtained by the FMM. The results are in good agreement with the experimental data. 
}
\label{fig1}
\end{figure}

The fabrication of the perovskite PCS (see Fig.~\ref{fig1}a) consists in the two steps: the synthesis of MAPbBr$_3$ thin film followed by the nanoimprint lithorgaphy. 

First, thin MAPbBr$_3$ film is synthesised by the solvent engineering method \cite{jeon2014solvent}. Perovskite solution is prepared in a nitrogen dry box by mixing of 56.0~mg of  methylammonium bromide (MABr, GreatCell Solar) and 183.5~mg of Lead(II) bromide (PbBr$_2$, TCI). Salts are dissolved in 1~ml 3:1 DMF:DMSO solvent mixture. The resulting solution with a molarity of 0.5M is stirred for 1 day at 27 C$^{\circ}$. Before a film synthesis, glass substrates (12.5$\times$12.5 mm) are cleaned by sonication in the deionized water, acetone and 2-propanol. In order to achieve high adhesion, the substrates are then cleaned in an oxygen plasma cleaner. The fabrication of a thin perovskite film is performed in a nitrogen dry box using spin-coating method. 30 uL of perovskite solution is deposited on the substrate and spinned at 3000~rpm for 40~sec. At the 25th second, 300~$\mu$L of toluene is dripped on the top of the rotating substrate. After the spinning, MAPbBr$_3$ film in the intermediate phase without thermal annealing is taken from the glovebox for further structuring using nanoimprint lithography method \cite{tiguntseva2019enhanced}.

As a mold for the nanoimrpint, we use a one-dimesional periodic structure with a period of 750~nm, ridges height of 100~nm and ridge width of 520~nm. The mold is cleaned in methanol and deionized water and then dried before the imprint. The imprint process is carried out under 4~MPa pressure for 10~minutes, then the mold is removed. Finally, the imprinted perovskite sample is annealed at 70 C$^{\circ}$ for 10~minutes. After the nanoimprint process, the perovskite nanograting is formed with the profile inverted with respect to the one of the mold employed.

In order to characterize the structural properties of the resulting perovskite PCS, we use the atomic force microscopy (AFM) and scanning electron microscopy (SEM) methods Fig.~\ref{fig1}b-d. The obtained data confirm the high quality of imprinted periodic nanograting and allow to precisely determine its parameters, in particular the period of the structure of 750~nm, ridge height of 45~nm, and ridge width of 200~nm (Fig.~\ref{fig1}c-d). Note that rare pinholes which can be seen in Fig.~\ref{fig1}d play a negligible role in the resulting optical quality of the sample. With extracted grating parameters and accounting for the measured refractive index dispersion of MAPbBr$_3$ thin films \cite{alias2016optical}, we calculate angle-resolved reflectance spectrum (Fig.~\ref{fig1}f) by Fourier modal method (FMM) \cite{li1997new}. 

In order to experimentally confirm the realization of the strong coupling regime in our sample at different temperatures, we use the angle-resolved spectroscopy methods. For this purpose, the back focal plane (BFP) of the objective lens Mitutoyo NIR x50 with N.A. equal to 0.65 is imaged to a slit spectrometer coupled to the liquid-nitrogen-cooled imaging CCD camera (Princeton Instruments SP2500+PyLoN). A halogen lamp is employed for the white light illumination in angle-resolved reflectance measurements. The white light homogeneously fills BFP, ensuring light incidence at the sample from the whole available range of the angles. The reflected light passes through a linear polarizer with electric field parallel to the grating ridges such that only TE modes are analyzed. Light is then dispersed by a monochromator and detected by the imaging CCD camera. As a result, we obtain the angular dependence of the reflectance spectra. The map shown in Fig.~\ref{fig1}e is obtained at room temperature and correlates well with the simulated one presented in Fig~\ref{fig1}f, which confirms the sufficient accuracy of the structural and optical parameters used in the simulation. In both Figs.~\ref{fig1}e,f one can observe the curvature of the optical mode dispersion in the vicinity of the exciton resonance around 2.31~eV, which is readily a clear signature of strong exciton-photon coupling and  polaritonic behavior. Due to the high absorption in the range of photon energies above the exciton level, no upper polariton branch can be observed in the experiment.

\begin{figure}
\centering
\center{\includegraphics[width=0.9\linewidth]{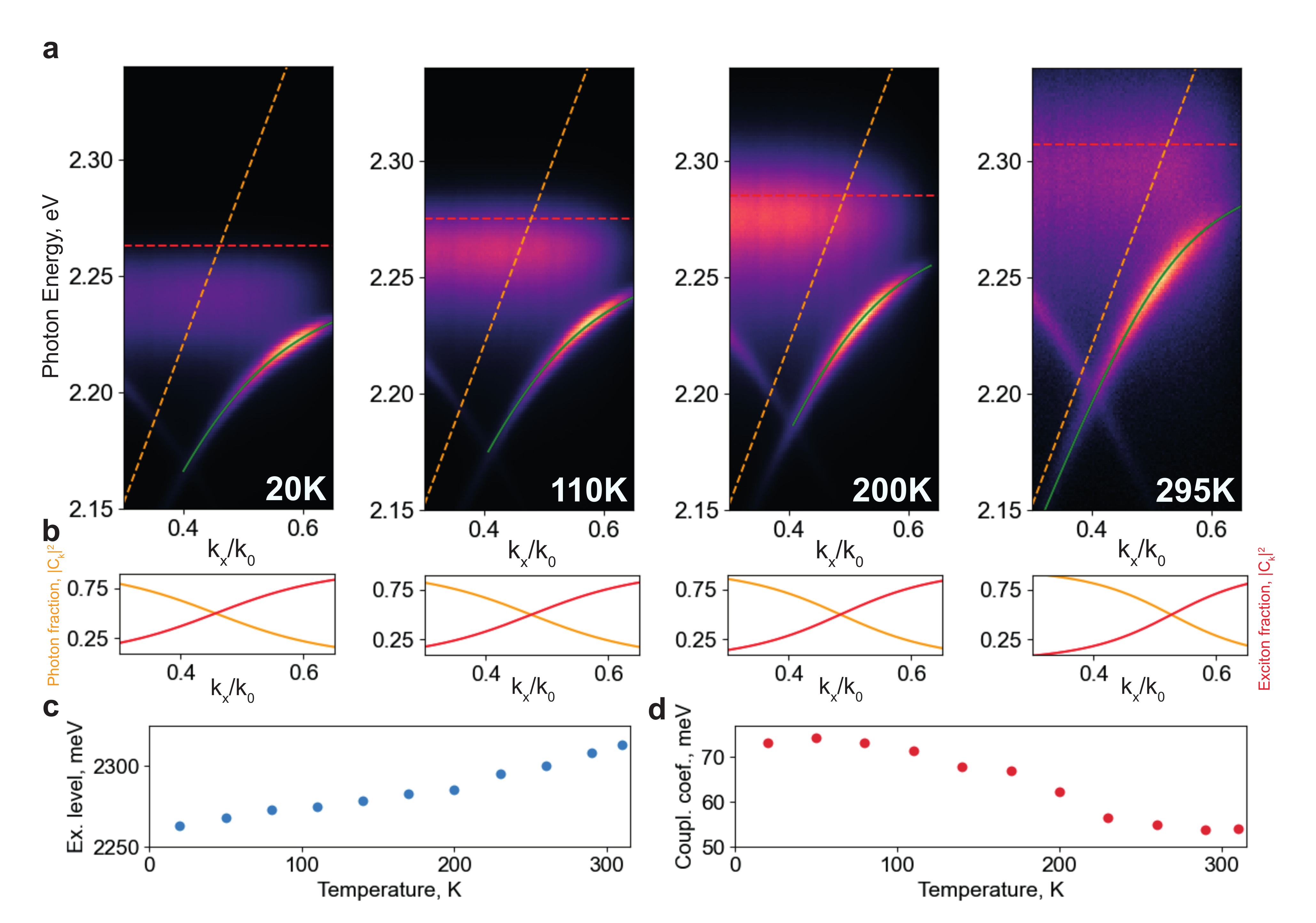}}
\caption{(a) Experimentally measured angle-resolved PL spectra at different temperatures under non-resonant pump. The color corresponds to the intensity of the emitted light. The data contain the background emission over all k$_x$/k$_0$ and emission of the leaky modes in the strong coupling regime. The dashed orange lines correspond to the uncoupled cavity photon mode, red dashed lines correspond to the exciton levels and the green solid lines describe the polariton modes, obtained by coupled oscillator model optimization. The observed shift of the lower polariton mode with temperature is attributed with the shift of the excitonic level. (b) Extracted Hopfield coefficients as  functions of  k$_x$/k$_0$ for given temperatures, estimated from the results of coupled oscillator model optimization. Orange and red lines correspond to the photon ($|C_k|^2$) and exciton ($|X_k|^2$) fractions respectively. (c,d) The extracted values of the excitonic energy $E_X$ in MAPbBr$_3$ PCS and light-matter coupling coefficient $g_0$ as functions of the temperature. The redshift of the exciton level with lowering the temperature is attributed to the polaron effects\cite{soufiani2015polaronic}. The rise of $g_0$ at low temperatures due to the increasing of the exciton fraction and lowering of the non-radiative losses.}
\label{fig2}
\end{figure}

\section*{Temperature-dependent strong light-matter coupling}

To study the light-matter coupling in our sample we perform angle-resolved photoluminescence (PL) measurements using the setup described above. To excite the sample non-resonantly, we use the femtosecond (fs) laser (Pharos, Light Conversion) coupled with the broad-bandwidth optical parametric amplifier (Orpheus-F, Light Conversion) at the wavelength of 490 nm (2.53~eV), 220 fs pulse duration and 100 kHz repetition rate. The sample is mounted in the ultra-low-vibration closed-cycle helium cryostat (Advanced Research Systems) and is maintained at a controllable temperature in the range of 7-300K.

Fig.~\ref{fig2}a shows the measured angle-resolved PL spectra for the temperatures varying from 20~K to 295K.  The spectral maps reveal two features: the narrow-band angle-dependent polariton mode and an uncoupled dispersionless emission in the vicinity of the exciton resonance. With increase of the temperature, we clearly observe the pronounced blueshift  of both the uncoupled PL peak and the polariton mode. 

In order to analyze carefully the evolution of exciton-polaritons with increase of the temperature, we fit the polariton dispersion using two-coupled oscillator model\cite{hopfield1958theory}:
\begin{equation}
    \widetilde{E_{LP}}(k_x) = \frac{\widetilde{E_X} + \widetilde{E_C}(k_x)}{2} - \frac{1}{2}\sqrt{(\widetilde{E_X} - \widetilde{E_C}(k_x))^2 + 4g_0^2},
\label{twocouplmod}
\end{equation}
where $\widetilde{E_X} = E_X - i \gamma_X$ is a complex exciton energy and $\widetilde{E_C}(k_x) = E_C(k_x) - i \gamma_C(k_x)$ is a complex photonic cavity mode energy, with imaginary parts $\gamma_X$ and $\gamma_C$ being the characteristic broadenings of excitonic and photonic modes, respectively, and $k_x$ is the in-plane component of the wavevector of light. 
Within the light cone the excitonic dispersion can be safely neglected.
$g_0$ is the strength of the light-matter coupling, which together with the broadenings of the modes define the Rabi splitting as follows:
\begin{equation}
    \Omega_{R} = \sqrt{4 g_0^2 - (\gamma_C - \gamma_X)^2}
\label{RabiSplit}
\end{equation}

Excitonic and photonic fractions in a lower polariton ($|X_k|^2$ and $|C_k|^2$ respectively), known as Hopfield coefficients, are shown in Fig~\ref{fig2}b. They depend on a wavevector and can be extracted using the fitting parameters of the dispersions using the following relations:
\begin{equation}
    |X_k|^2 = \frac{1}{2} \left(1- \frac{E_C(k_x) - E_X}{\sqrt{(E_C(k_x) - E_X)^2 + 4 g_0^2}} \right); \; 
    |C_k|^2 = 1 - |X_k|^2.
\label{HopfCoeff}
\end{equation}

The uncoupled cavity mode (orange lines in Fig.~\ref{fig2}a) is obtained by a linear extrapolation of the leaky mode dispersion from the spectral region far from the excitonic resonance to the whole spectral range. The polariton dispersion $\widetilde{E_{LP}}(k)$ is extracted from the measured angle-resolved PL spectrum by the fitting the mode resonance at each in-plane wavevector k$_x$/k$_0$ by the Lorentz peak function. 
Here, $k_0$ is a wavevector of light in free space defined as $k_0 = E / (\hbar * c)$, where $E$ is photon energy, $c$ is the speed of light.
It should be noted, that in the recent works, some uncertainty in the determination of exciton resonance in MAPbBr$_3$ depending on the method was reported \cite{soufiani2015polaronic, shi2020exciton} and we therefore choose $\widetilde{E_X}$ and $\Omega_0$ in Eq. \eqref{twocouplmod} as optimization parameters for each temperature in our fitting procedure. 

Fig.~\ref{fig2}d shows the temperature dependence of the coupling coefficient $g_0$. The fitting procedure also allows to extract the values of $\gamma_C\approx 20$~meV and $\gamma_X \leq \gamma_X^{RT} \approx 12$~meV, where $\gamma_X^{RT}$ is the room temperature exciton linewidth [see {\color{blue}Supplementary Material (SM)}]. Taking into account the strong light-matter coupling criteria $g_0 > |\gamma_C - \gamma_x|/2$ and $\Omega_{R} > (\gamma_C+\gamma_X)/2$, we confirm that in our system exciton-polaritons remain stable up to the room temperatures.  In particular, at room temperature we get the value $g_0 = 54$~meV, which exceeds the value of $g_0 = 48$~meV reported previously for the geometry with a vertical Bragg cavity\cite{bouteyre2019room}.  This enhancement can be attributed to a better field localization in the PCS of our sample .

The temperature dependence of the position of the excitonic resonance (Fig.~\ref{fig2}c) extracted from the two-coupled oscillator model is in good agreement with the previous works \cite{shi2020exciton,soufiani2015polaronic}. Note that in case of the fitting of an uncoupled resonance from the reflectance measurements \cite{soufiani2015polaronic}, one should account for the inhomogeneous broadening of an excitonic line, while in the case of the strong exciton-photon coupling, this broadening is lifted due to the effect of motional narrowing\cite{whittaker1996motional}. The pronounced temperature dependence of the position of an excitonic level is the signature of strong polaronic effects in the linear optical response \cite{soufiani2015polaronic}. Below we demonstrate that these effects play crucial role in the nonlinear optical response in MAPbBr$_3$ as well.

\section*{Nonlinear polariton blueshift under resonant pump}

\begin{figure}
\centering
\center{\includegraphics[width=0.8\linewidth]{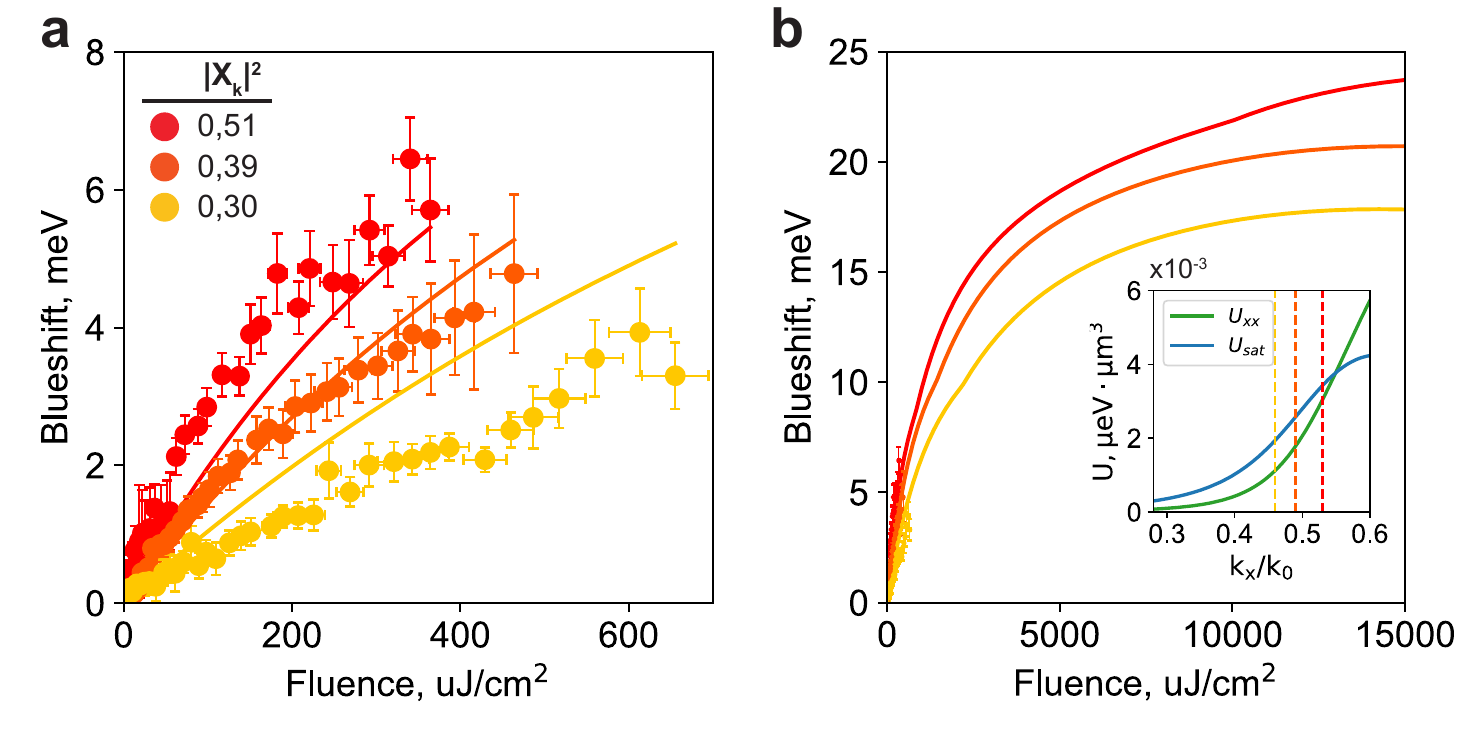}}
\caption{ (a) Extracted polariton blueshift as a function of incident pump fluence of the femtosecond laser excitation at room temperature.
The thin solid lines are the results of theoretical modeling.
Inset table shows coefficients $|X_k|^2$ for excitonic fractions in polaritons.
In the experiment, the observable blueshifts are limited by the spectral linewidth of the pulse. 
We therefore perform the numerical simulations for the same set of parameters within the extended range of  pump fluences reaching the blueshift saturation, as shown in panel (b).
Inset: Wavevector dependence of polariton nonlinearity rates associated with exciton-exciton Coulomb interaction (blue curve) and the quench of Rabi splitting due to the phase space filling (green curve). 
The vertical dashed lines of corresponding color indicate  the experimental wavevectors shown in panels (a), (b). 
The nonlinearity associated with the quench of Rabi splitting is comparable with exciton-exciton interactions, which stems from the very large value of bare light-matter coupling $g_0 = 54$~meV.
}
\label{fig3}
\end{figure}

In order to study the role of the polaronic effects in polariton nonlinearity, we first perform resonant pump-dependent reflectivity measurements. 
The sample is excited by 220 fs pulses with tunable central wavelength.
The angle of the incidence is controlled by the focusing of the laser beam to the back focal plane of an objective lens. 
We select several angles of pump incidence  for near-resonant excitation of polaritons with different exciton fractions defined by Hopfield coefficient $|X_k|^2$, eq.~(\ref{HopfCoeff}). 
We measure the reflectance spectra within the spectral range of the pump fs laser and extract the polariton resonance peak parameters by fitting it with Fano lineshape function. 
With increase of the pump fluence, we observe the nonlinear blueshift of the lower polariton line. 
The measured blueshifts as functions of the incident fluence for different wavevectors (corresponding to different values of Hopfield coefficients) are shown in Fig~\ref{fig3}a with circles of the different colors. 
Naturally, higher polariton blueshifts are observed for higher excitonic fractions $|X_k|^2$.
For the highest exciton fraction we achieve the maximum measured blueshift of 6.4 meV. In our case this quantity is experimentally limited by the spectral linewidth of the pump laser, and can be expected to increase even further.

We now proceed with the theoretical model for the quantitative description of the polaron-assisted polariton nonlinearity, which will allow us to describe the presented experimental data.
The Schr\"{o}dinger equation for an excitonic state in the center of mass coordinates reads:
\begin{equation}
    \label{eq:Schro}
    \left[-\frac{\hbar^2 \nabla^2}{2\mu^*} - V(\vec{r}) \right] \psi(\vec{r}) = -E_b \psi(\vec{r}),
\end{equation}
where $\mu^*=m_e^* m_h^*/ (m_e^*+m_h^*)$ is the exciton reduced mass, with $m_e^*$ and $m_h^*$ being the masses of electrons and holes. In halide perovskites, the strong coupling between the charge carriers and longitudinal optical (LO) phonon modes leads the formation of polarons and corresponding  renormalization of the charge carrier effective masses, which can be represented as\cite{masharin2022polaron},
\begin{equation}
m_{e[h]}^* = m_{e[h]} \left(1+ \alpha_{e[h]}/6 \right),
\end{equation}
where
\begin{equation}
    \alpha_{e[h]} = \frac{e^2}{4\pi\varepsilon_0} \frac{1}{\hbar \varepsilon^*} \sqrt{\frac{m_{e[h]}}{2E_{LO}}}
\end{equation}
is the dimensionless Fröhlich coupling constant. In the above expression $E_{LO}$ is the energy of LO phonon, 
$e$ is the elementary charge, $\varepsilon_0$ is the vacuum permittivity, and
\begin{equation}
    \frac{1}{\varepsilon^*} = \frac{1 }{\varepsilon_{\infty}}
    - \frac{1}{\varepsilon_s},
\end{equation}
with $\varepsilon_s$, $\varepsilon_\infty$ being static and high frequency dielectric constants, respectively.

Besides effective mass renormalization, polaronic effects strongly affect the interaction between the charge carriers, which within the Pollmann-Büttner approach can be represented as \cite{pollmann1977effective,baranowski2020excitons,menendez2015nonhydrogenic}
\begin{equation}
    \label{eq:Vr}
    V(r) =  \frac{e^2}{4\pi \varepsilon_0 r} \left[ \frac{1}{\varepsilon_s } +\frac{1}{\varepsilon^* } 
    \left( \frac{m_h}{\Delta m} e^{-\frac{r}{l_h}} - \frac{m_e}{\Delta m} e^{-\frac{r}{l_e}}\right) \right],
\end{equation}
where $\Delta m = m_h - m_e$, and $l_{e[h]} = \hbar /\sqrt{E_{LO} m_{e[h]} } $
are the polaron radii. 

For numerical calculations, we use the parameters from the Refs. \cite{baranowski2020excitons,soufiani2015polaronic}. 
We set $\varepsilon_{\infty} = 4.4$,
$\varepsilon_{s} = 25.5$, $E_{LO}=18$ meV, $m_e = 0.21 m_0$, $m_h = 0.25 m_0$, where $m_0$ is the free electron mass. 
We numerically calculate the eigenvalues of Eq.~\eqref{eq:Schro}, and obtain the excitonic binding energy $E_b = 30.5$ meV, which is in line with previous reports ~\cite{soufiani2015polaronic,shi2020exciton}.
The obtained wave function of the ground state of an exciton  is well fitted with the hydrogen-type expression
$\psi = \frac{1}{\sqrt{\pi a^3}} e^{-r/a_B}$, where $a_b \approx 1.76$ nm is the excitonic Bohr radius.

The high value of the binding energy makes excitons stable at room temperatures.
Note, that the polaronic renormalization of the effective masses and the interaction potential leads to the  violation of conventional hydrogenic Rydberg scaling of exciton energy levels and the Bohr radius \cite{menendez2015nonhydrogenic}.
It also modifies substantially the relations between the exciton Bohr radius (which defines the maximally reachable exciton density, corresponding to the Mott transition) and the nonlinear interaction rates in a way favorable for enhancing the fluence-dependent blueshift \cite{masharin2022polaron}.

To demonstrate that, we calculated the matrix elements of the exciton-exciton interaction, and Pauli factors responsible for the quenching of the Rabi splitting which govern the nonlinear optical response.
The resulting blueshift of the lower polariton mode can be represented in form of the series in terms of the polariton density $n_{Lk}$ \cite{combescot2008many,emmanuele2020highly}.
Keeping only the terms linear and quadratic in $n_{Lk}$ one gets:
\begin{align}
    \label{eq:Epoltot}
    \Delta E_{LP}(k_x, n_{Lk}) \approx U (k_x)  n_{Lk} -  U_2 (k_x)  n_{Lk}^2 +\hat{O} (n_{Lk}^3),
\end{align}
where the expansion coefficients read \cite{emmanuele2020highly,masharin2022polaron}:
\begin{align}
    \label{eq:U}
    U (k_x) \approx 
    \frac{V_{XX}}{2} |X_k|^4 
    + g_0 s |X_k|^2 (X_k^* C_k + X_k C_k^*)
    := U_{XX}(k_x) + U_{\rm sat} (k_x),
\end{align}
\begin{align}
    \label{eq:U2}
    U_2 (k_x) \approx 
    |V_{XX2}| |X_k|^6 
    + g_0 |s_2| |X_k|^4 (X_k^* C_k + X_k C_k^*)
    := U_{XX2}(k_x) + U_{\rm sat 2} (k_x),
\end{align}
Each of the terms represents the sum of the contributions stemming from the exciton-exciton scattering and renormalization of the Rabi spltiinng, the latter being governed by the saturation rates $s$, $s_2$. The explicit expressions for the corresponding parameters are given in \blue{SM}.
Due to the hydrogen-type shape of the excitonic wave function, the saturation factors can be computed analytically, as $s = 7\pi a_B^3 $, $s_2 = - 253 \pi^2 a_B^6 /4$. 
The calculated values of the exciton-exciton Coulomb scattering matrix elements are $V_{XX} = 0.022$ $\mu$eV$\cdot \mu$m$^3$, $V_{XX2} = -1.1 \cdot 10^{-9}$ $\mu$eV$\cdot \mu$m$^6$.

In order to determine the dependence of the polariton  density on the intensity of a resonant pump in nonlinear experiments, we use the input-output formalism. 
In the mean field approximation, the polariton density can be written as the square of the absolute cvalue of the coherent lower polariton field, $n_{Lk} = |\bar{p}_k|^2$, where 
$\bar{p}_k = \langle \hat{p}_{Lk} \rangle$, with $\hat{p}_{Lk}$ being the lower polariton annihilation operator.

The dynamic equation for $\bar{p}_k$ reads:
\begin{align}
    \label{eq:pdyn}
    \hbar \dot{\bar{p}}_k = & - i E_{LP} (k_x) \bar{p}_k
    - \frac{ \gamma_{L}(k_x) + \gamma^\prime}{2} \bar{p}_k
    -|X_k|^4 \gamma_2 |\bar{p}_k|^2 \bar{p}_k  
    -  2 i U(k_x)|\bar{p}_k|^2 \bar{p}_k    
    +  3 i U_2(k_x) |\bar{p}_k|^4 \bar{p}_k 
    + 
    \sqrt{\frac{\gamma_L (k_x)}{2} } a_0 e^{-t^2/(2\tau_p^2)} e^{-i \omega_Lt},
\end{align}
where $\gamma_L (k_x)$ [$\gamma^\prime$] are radiative [non-radiative] decay rates of the lower polariton branch, $\tau_p$ is the pulse duration, $\omega_L$ is the pump frequency.
$a_0$ is the
square root of the number of photons passing through the structure per unit time per unit
area, which is related to the peak incident power density as $a_0 = \sqrt{F/(\omega_{LP} \tau_p L_C )}$, with $F$ being the pump fluence, and $L_C =0.3$ $\mu$m the thickness of the sample. 
We found the ratio $\gamma_L (k_x) / [\gamma_L (k_x) + \gamma^\prime] \approx 0.15$ in a wide range of the values of $k_x$.
The parameter $\gamma_2 $ is the exciton-exciton annihilation rate, defined by collisional broadening \cite{ciuti1998role}. 
We numerically simulate the Eq.~\eqref{eq:pdyn} and assume that the optical response is collected at maximal density, $n_{Lk} = n_{Lk}(t)|_{\rm max}$.
The exciton-exciton annihilation rate is treated as fitting parameter and chosen as  $\gamma_2 = 1.5 V_{XX}$.

The developed theory allowed us to fit the experimental values of the blueshift as functions of the fluence. 
The contributions of both exciton-exciton interaction and absorption saturation were accounted for and are both determined by excitonic fraction in a lower polariton, which varies with the in-plane wave vector $k_x$. 
This fact explains the wave-vector dependence of the nonlinear blueshift, shown in Fig.~\ref{fig3}. 

The trend towards the saturation of the blueshift with increase of the fluence, clearly visible in Fig.~\ref{fig3}b can stem from the quadratic in density terms, which have the sign opposite to the linear terms and thus describe the redshift contribution [see Eqs.\eqref{eq:Epoltot}, \eqref{eq:U2}].
However, our estimation showed that in such mechanism the blueshift saturation effects should become visible only for the fluences $F> 1000 $ $\mu$J/cm$^2$. 

We therefore attribute the sublinear fluence dependence of polariton blueshift in our case shown in Fig.~\ref{fig3}a primarily to the exciton-exciton annihilation, which limits the increase of the polariton density with pump fluence.
It was found earlier and confirmed by our experiments, that MAPbBr$_3$ films can support further increase of fluence up to $F\sim 10^4$ $\mu$J/cm$^2$ without irreversible changes in a sample  \cite{liu2017organic,wang2019flexible}, and allow to reach up to  $10^{19}$ cm$^{-3}$ of polariton density \cite{richter2016enhancing}. 
In our experiment, the maximum observable mode blueshift of $\approx 6$~meV is limited by the spectral linewidth of femtosecond laser radiation ($\approx 14$~meV). For larger blueshifts, the overlap between the spectral profile of the polariton mode (with the linewidth of $\approx 15$~meV) and the accessible spectral range defined by the spectrum of the pulse becomes insufficient, which makes the extraction of the blueshifted mode position inaccurate.
We therefore theoretically simulate this hypothetical regime and find that one can expect blueshift as large as 19 meV [see Fig.~\ref{fig3}b], similar to that observed in Ref. \cite{masharin2022polaron} for MAPbI$_3$ at cryogenic temperatures. 

The relative impact of the the two components of nonlinearity is presented in the inset of Fig.~\ref{fig3}b.
Notably, the contribution stemming from the quench of the Rabi splitting is quite large and comparable with the contribution coming from exciton-exciton scattering. 
This situation is quite unusual and is due to the large value of light-matter coupling $g_0 \sim 50$ meV.

\section*{Conclusion}

We demonstrated the robust polariton nonlinear response in MAPbBr$_3$ hybrid perovskites at room temperature. 
For this purpose, using the nanoimprint lithography, we have fabricated MAPbBr$_3$-based photonic crystal slab cavity supporting strong exciton-photon coupling in a wide temperature range. By fitting the polariton dispersions we demonstrated the substantial variation of the exciton transition energy with temperature, which is a fingerprint of the polaron effects in the studied system. 
These effects are responsible for the stability of the excitons at room temperatures and the enhancement of the optical nonlinearity, detected as the blueshift of the lower polariton line under resonant femtosecond pump. 
We developed the corresponding theoretical model, and demonstrated that the contribution associated with the phase space filling effects can be comparable and even exceed the one provided by exciton-exciton scattering. Predicted values of the blueshift achievable in MAPbBr$_3$ at room-temperature exceed 20 meV, which can provide good means for the realization of high-contrast optically-induced polaritonic potentials for study of the interacting polariton fluids, and their applications in optical analogue simulators.

\bibliography{sample}

\section*{Acknowledgements}
The experimental part of this work was funded by Russian Science Foundation, grant \#21-12-00218. The work by A.K.S. was supported by Mercur Foundation (Grant Pe-2019-0022) and TU Dortmund core funds. 
V.S. acknowledges the support of “Basis” Foundation (Project No. 22-1-3-43-1). I.A.S. acknowledges  the support of the Icelandic Research Fund (Rannis), project No. 163082-051. 

\end{document}